\documentclass[aps,prl,twocolumn,groupedaddress]{revtex4} 
\usepackage{epsfig,amsopn}
\begin{document}

\title{Coherent manipulation of electronic states in a double quantum dot}
\author{T. Hayashi$^{1}$}
\author{T. Fujisawa$^{1}$}
\author{H. D. Cheong$^{2}$}
\author{Y. H. Jeong$^{3}$}
\author{Y. Hirayama$^{1,4}$}
\affiliation{$^{1}$NTT Basic Research Laboratories, NTT Corporation, 3-1
Morinosato-Wakamiya, Atsugi, 243-0198, Japan}
\affiliation{$^{2}$Keimyung University, Shindang-Dong, Dalseo-gu, Daegu, Korea}
\affiliation{$^{3}$Pohang University of Science and Technology, Pohang, Kyungpook, Korea}
\affiliation{$^{4}$CREST, 4-1-8 Honmachi, Kawaguchi, 331-0012, Japan}

\begin{abstract}
We investigate coherent time-evolution of charge states (pseudo-spin qubit)
in a semiconductor double quantum dot. This fully-tunable qubit is
manipulated with a high-speed voltage pulse that controls the energy and
decoherence of the system. Coherent oscillations of the qubit are observed
for several combinations of many-body ground and excited states of the
quantum dots. Possible decoherence mechanisms in the present device are also
discussed.
\end{abstract}

\date{\today}
\maketitle

Initiated by various experiments on atomic systems, studies on coherent
dynamics have been extended to small-scale quantum computers \cite%
{BookNielsenChuang}. Nano-fabrication technology now allows us to design
artificial atoms (quantum dots) and molecules (coupled quantum dots), in
which atomic (molecular)- like electronic states can be controlled with
external voltages \cite{TaruchaAAPRL,OosterkampNature,FujisawaScience}. 
Coherent manipulation of the electronic system in quantum dots and a clear
understanding of decoherence in practical structures are crucial for future
applications of quantum nanostructures to quantum information technology.

In this Letter, we describe the coherent manipulation of charge states, in
which an excess electron occupies the left dot or the right dot of a double
quantum dot (DQD). The coherent oscillations between the two charge states
are produced by applying a rectangular voltage pulse to an electrode.
Although this scheme is analogous to experiments on a superconducting island %
\cite{NakamuraNature}, our qubit is effectively isolated from the electrodes
during the manipulation, while it is influenced by strong decoherence during
the initialization due to the coupling with the electrodes. This controlled
decoherence provides an efficient initialization scheme.

We consider a DQD consisting of left and right dots connected through an
interdot tunneling barrier. The left (right) dot is weakly coupled to the
source (drain) electrode via a tunneling barrier [see Fig. 1(a)]. The
conductance through the device is strongly influenced by the on-site and
interdot Coulomb interactions \cite{WilfredRMP}. In the weak-coupling regime
at a small source-drain voltage, $V_{sd}$, a finite current is only observed
at the triple points, where tunneling processes through the three tunneling
barriers are allowed. Under an appropriate condition where only the interdot
tunneling is allowed, Coulomb interactions effectively isolate the DQD from
the source and drain electrodes. In this case, we can consider two charge
states, in which an excess electron occupies the left dot ($|L\rangle $) or
the right dot ($|R\rangle $) with electrochemical potentials $E_{L}$ and $%
E_{R}$, respectively. In practice, each charge state involves (many-body)
ground and excited states. When the two specific states are energetically
close to each other and the excitation to other states can be neglected, the
system can be approximated as a two-level system (qubit). It is
characterized by the energy offset, $\varepsilon \equiv E_{R}-E_{L}$, and
the interdot tunneling, which gives an anti-crossing energy, $\Delta $ \cite%
{OosterkampNature}. The effective Hamiltonian is%
\begin{equation}
H=\frac{1}{2}\varepsilon (t)\sigma _{z}+\frac{1}{2}\Delta \sigma _{x},
\label{EqHamiltonian}
\end{equation}%
where $\sigma _{x}$ and $\sigma _{z}$ are the Pauli matrices for pseudo-spin
bases of $|L\rangle $ and $|R\rangle $. When $E_{L}$ and $E_{R}$ of the
localized states are crossed by changing $V_{sd}$, for instance, as shown by
dashed lines in Fig. 1(b), the eigenenergies, $E_{b}$ and $E_{a}$, for
bonding and anti-bonding states respectively, show anti-crossing as shown by
solid lines. The coherent oscillation of the system is expected with the
angular frequency given by $\Omega =\sqrt{\varepsilon ^{2}+\Delta ^{2}}%
/\hbar $.

\begin{figure}[t]
\epsfxsize=3.3in \epsfbox{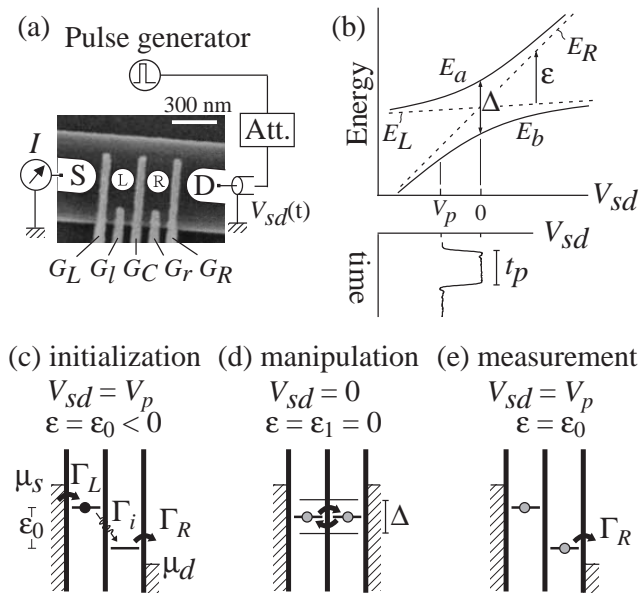} 
\caption{(a) Schematic measurement circuit combined with a scanning electron
microscope image of the sample. Etching (upper and lower dark regions) and
negatively biased gate electrodes ($G_{L}$, $G_{l}$, $G_{C}$, $G_{r}$, and $%
G_{R}$) define a double quantum dot (L and R) between the source (S) and
drain (D). (b) Energy levels of the bonding ($E_{b}$) and anti-bonding ($%
E_{a}$) states, which are the eigenstates during the manipulation, and
localized states ($E_{L}$ and $E_{R}$) during initialization. A typical
condition of $\protect\varepsilon _{1}=0$, where $E_{L}$ and $E_{R}$ cross
at $V_{sd}=0$, is shown. A typical pulsed voltage $V_{sd}(t)$ is shown at
the bottom. (c) - (e) Energy diagrams of the DQD for $\protect\varepsilon %
_{1}=0$ during (c) initialization, (d) coherent oscillation, and (e)
measurement process.}
\end{figure}

The DQDs (samples I and II with almost identical dimensions) used for this
work are defined in a GaAs/AlGaAs heterostructure containing a
two-dimensional electron gas, as shown in Fig. 1(a). The experiments were
performed in a magnetic field of 0.5 T at lattice temperature $%
T_{lat}\lesssim $ 20 mK, unless otherwise noted. The effective electron
temperature, however, remained at $T_{elec}\sim $ 100 mK. Each dot in both
samples contains about 25 electrons and has an on-site charging energy $%
E_{c}\sim $ 1.3 meV. The interdot electrostatic coupling energy is $U\sim $
200 $\mu $eV. Figure 2(a) shows the current spectrum $I$ of sample I when
the voltage, $V_{R}$, on the right gate [$G_{R}$ in Fig 1(a)] is swept at a
large source-drain voltage $V_{sd}=$ 650 $\mu $V. Each dot contains several
energy states in the transport window of width $eV_{sd}$, and resonant
tunneling between them is clearly resolved as current peaks, two of which
(resonances $\alpha $ and $\beta $) are shown in Fig. 2(a). Resonance $%
\alpha $ ($\beta $) is probably associated with the ground state of the left
dot and the first (second) excited state of the right dot. In the vicinity
of each peak, a two-level system (qubit) can be defined by only taking into
account a single discrete state in each dot, $|L\rangle $ and $|R\rangle $.
The qubit parameters, $\varepsilon $ and $\Delta $, and tunneling rates, $%
\Gamma _{L}$ and $\Gamma _{R}$, respectively for left and right barriers,
can be controlled independently by external gate voltages, and are
determined from the elastic current spectra \cite%
{OosterkampNature,FujisawaScience,WilfredRMP}.

In order to manipulate the qubit, a rectangular voltage pulse is applied to
the drain electrode.\ This switches the source-drain bias voltage $V_{sd}$
between $V_{p}=$ 650 $\mu $V, at which the tunneling between the DQD and the
electrodes is allowed, and zero, at which the DQD is effectively isolated
from the electrode due to Coulomb interactions. At the same time, due to the
electrostatic coupling between the QDs and electrodes, the pulse also
switches the energy offset between $\varepsilon =\varepsilon _{0}$ at $%
V_{sd}=V_{p}$ and $\varepsilon _{1}$ at $V_{sd}=0$ ($\varepsilon
_{1}-\varepsilon _{0}\sim $ 30 $\mu $eV), as shown for $\varepsilon _{1}=0$
in Fig. 1(b). We designed the pulse sequence for initialization, coherent
manipulation, and measurement in the following way.

For initialization, a relatively large source-drain voltage, $V_{sd}=V_{p}$,
was applied under appropriate gate voltages, so that $E_{L}$ and $E_{R}$ are
in between the electrochemical potentials of the source and drain
electrodes, $\mu _{S}$ and $\mu _{D}$ ($\mu _{S}>E_{L},E_{R}>\mu _{D}=\mu
_{S}-eV_{p}$). For example, in the off-resonance condition ($\varepsilon
=\varepsilon _{0}\lesssim -\Delta $) as shown in Fig. 1(c), electron-phonon
interaction provides finite inelastic tunneling, whose rate is $\Gamma _{i}$%
, between the two states \cite{FujisawaScience}. We adjusted $\Gamma _{L}$
and $\Gamma _{R}$ to make them sufficiently larger than $\Gamma _{i}$ so
that the current would be limited by the inelastic tunneling between the
dots. This sequential tunneling process accumulates an excess electron in
the left dot, providing the initial state $|L\rangle $. Note that this
initialization works even in the resonance condition ($\varepsilon _{0}=0$)
when $\hbar \Gamma _{L}$ and $\hbar \Gamma _{R}$ are greater than $\Delta $.
Significant decoherence from the dissipative tunneling processes holds the
system in the localized state $|L\rangle $ rather than in the delocalized
states.

For coherent manipulation, we non-adiabatically change $V_{sd}$ to zero,
which shifts the energy offset to $\varepsilon =\varepsilon _{1}$. A typical
energy diagram for $\varepsilon _{1}\sim $ 0 is shown in Fig. 1(d). In this
case, the inter-dot electrostatic coupling prevents the electron tunneling
into and out of the DQD by any first-order tunneling process, and negligible
current flows through the DQD. Hence the system is well approximated by Eq. %
\ref{EqHamiltonian}. The system prepared in $|L\rangle $ goes back and forth
between $|L\rangle $ and $|R\rangle $ coherently. We maintain $V_{sd}=0$ for
the pulse length, $t_{p}=$ 80 - 2000 ps, during which the oscillation
continues.

Then, the large bias voltage is restored for the measurement [Fig. 1(e)].
The large tunneling rates ($\hbar \Gamma _{L},$ $\hbar \Gamma _{R}>\Delta $)
effectively stop the coherent manipulation, and thereby provide a strong
measurement. If the system ends up in $|R\rangle $ after the manipulation,
the electron tunnels out to the drain electrode and contributes to the
pumping current. The system goes back to the initial state $|L\rangle $
after waiting longer than $\Gamma _{L}^{-1}+$ $\Gamma _{R}^{-1}$. However,
no pumping current is expected for $|L\rangle $, which is already the
initial state. Hence, this pumping current depends on the probability of
finding the system in $|R\rangle $.

In practice, we repeatedly applied many pulses with a repetition frequency $%
f_{rep}$ = 100 MHz and measured the average dc current, $I$, which comprises
the coherent pumping current and inelastic current that flows during
initialization. In order to improve the signal-to-noise ratio, we employed a
lock-in amplifier technique to measure the pulse-induced current $I_{p}$ by
switching the pulse train on and off at a low modulation frequency of 100
Hz. We estimated the average number of pulse-induced tunneling electrons, $%
n_{p}=I_{p}/ef_{rep}$.

A color plot of $n_{p}$ as functions of $V_{R}$ and $t_{p}$ is shown in Fig.
2(b). Sweeping $V_{R}$ mainly shifts $E_{R}$ and changes the energies $%
\varepsilon _{0}$ and $\varepsilon _{1}$ simultaneously by keeping $%
\varepsilon _{1}-\varepsilon _{0}$ almost constant. A clear oscillation
pattern is observed in a wide range of $V_{R}$. Local-maxima of the
oscillation amplitude appeared for relatively long $t_{p}$ at gate voltages
indicated by long-dashed lines, where the two states must be resonant ($%
\varepsilon _{1}$ = 0) during manipulation. We confirmed that the
oscillation patterns in Fig. 2(b) are attributed to resonances $\alpha $
(clear oscillation) and $\beta $ (faint oscillation) from their $V_{p}$
dependence. The energy offsets $\varepsilon _{0,i}$ and $\varepsilon _{1,i}$
for resonance $i$ ($\alpha $ and $\beta $) are also shown in Fig. 2(b).

The oscillation pattern for resonance $\alpha $ shows that the amplitude and
period decrease as $\varepsilon _{1,\alpha }$ goes away from $\varepsilon
_{1,\alpha }=0$. The current amplitude is asymmetric about $\varepsilon
_{1,\alpha }=0$, and the oscillation continues until $\varepsilon _{1,\alpha
}\sim $ 40 $\mu $eV. These features are qualitatively consistent with a
calculation based on the time-dependent Schr\"{o}dinger equation and Eq. \ref%
{EqHamiltonian} using a time-dependent $\varepsilon (t)$ with a finite rise
time ($\sim $ 100 ps) of the pulse \cite{NakamuraNature}. The strong
oscillation in the range of $\varepsilon _{0,\alpha }<0<\varepsilon
_{1,\alpha }$ can be understood as an interference between coherent
time-evolution at $\varepsilon (t)\sim $ 0 during the finite rise time of
the pulse and that during the fall time of the pulse. It should be noted
that clear oscillation is seen even at $\varepsilon _{0,\alpha }=0$
(indicated by a black dotted line), where two localized states are resonant
during the initialization but off-resonant during the manipulation. This
feature is convincing evidence that there is strong decoherence during
initialization. The density matrix calculation for our initialization
condition gives the decoherence rate, $\hbar (\Gamma _{R}+\Gamma _{L})/2\sim 
$ 30 $\mu $eV \cite{StoofPRB}, which is greater than $\Delta =$ 9 $\mu $eV
for resonance $\alpha $. However, the Coulomb blockade effect eliminates
this decoherence during manipulation, as mentioned before. Therefore, we
presume that the oscillation at $\varepsilon _{0,\alpha }=0$ is induced by
the modulation of the decoherence rate. In contrast, the disappearance of
the oscillation at $\varepsilon _{0,\beta }=0$ (indicated by a white dotted
line) for resonance $\beta $ ($\Delta =$ 30 $\mu $eV) might arise from the
inefficient initialization that provides a statistical mixture of bonding
and antibonding states.

The qubit state can be manipulated arbitrarily. Ideally, the quarter period
oscillation at $\varepsilon _{1}=0$ corresponds to the $\pi /2$ pulse that
prepares a superposition state $\frac{1}{\sqrt{2}}(|L\rangle +i|R\rangle )$.
Leaving a state at $\varepsilon =\varepsilon _{2}\gg \Delta $ for a specific
time $t_{\phi }$ gives a phase shift $\varepsilon _{2}t_{\phi }/\hbar $
between $|L\rangle $ and $|R\rangle $. Therefore arbitrary states can be
prepared by tailoring the pulse waveform $\varepsilon (t)$ even at a
constant $\Delta $. The demonstration of phase-shift operations will be
published elsewhere \cite{PhaseGate}.

Figure 2(c) shows typical $n_{p}(t_{p})$ traces at $\varepsilon _{1}=0$ for
resonances $\alpha $ and $\beta $. The oscillation can be fitted well by an
exponential decay of the cosine function and a linearly decreasing term,%
\begin{equation}
n_{p}(t_{p})\simeq A-\frac{1}{2}B\exp (-t_{p}/T_{2})\cos (\Omega
t_{p})-\Gamma _{i}t_{p},  \label{EqNp}
\end{equation}%
except when $t_{p}\lesssim $ 100 ps (the rise time of the pulse). The last
term comes from the fact that the inelastic tunneling current is blocked
during the manipulation. Actually $\Gamma _{i}\sim $ (6 ns)$^{-1}$ obtained
for $\alpha $ from this fitting is consistent with the inelastic dc current,
which should be $e\Gamma _{i}$ in the absence of the pulse. The offset, $%
A\sim $ 0.6, and amplitude, $B\sim $ 0.3, of the oscillation for $\alpha $
are comparable to the ideal case ($A=$ 0.5 and $B=$ 1 at $\varepsilon _{1}=0$%
), although they are degraded by the finite rise time of the pulse and
non-ideal initialization/measurement processes. The oscillation frequency $%
\Omega $ and the decoherence time $T_{2}$ can be obtained from the fitting ($%
\Omega /2\pi \sim $ 2.3 GHz and $T_{2}\sim $ 1 ns for resonance $\alpha $ at 
$\varepsilon _{1}=0$).

\begin{figure}[t]
\epsfxsize=3.3in \epsfbox{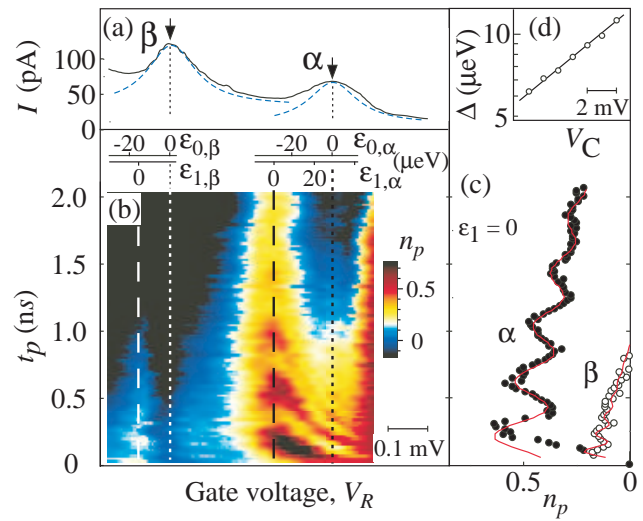} 
\caption{(Color) (a) Current profile, $I$ vs $V_{R}$, at constant $V_{sd}=$
650 $\protect\mu $V. Two resonant tunneling peaks, $\protect\alpha $ and $%
\protect\beta $, out of about six peaks in the transport window, are shown.
The ground state resonant peak (not shown) is located at about 0.5 mV to the
right of peak $\protect\alpha $. Lorentzian fitting (dashed lines) to peaks $%
\protect\alpha $ and $\protect\beta $ gives approximate parameters $\hbar
\Gamma _{L}\sim \hbar \Gamma _{R}$ $\sim $ 30 $\protect\mu $eV. (b) Color
plot of $n_{p}$ as a function of $V_{R}$ and $t_{p}$. The horizontal axis is
also shown in terms of $\protect\varepsilon _{0,i}$ and $\protect\varepsilon %
_{1,i}$ for resonance $i$ ($\protect\alpha $ or $\protect\beta $). (c) $%
n_{p}(t_{p})$ at $\protect\varepsilon _{1}=0$ (long-dashed lines in b) for
the resonance $\protect\alpha $ (solid circles) and $\protect\beta $ (open
circles). Lines are fitted to the data. (d) The coupling energy, $\Delta $,
determined from the oscillation frequency, when the gate voltage on $G_{C}$
is changed. The line is a guide for the eye.}
\end{figure}

We estimate how the decoherence rate $T_{2}^{-1}$ depends on the energy
offset $\varepsilon _{1}$ [Fig. 3(a)], coupling energy $\Delta $ [Fig.
3(b)], and the lattice temperature $T_{lat}$ [Fig. 3(c)]. Here $\Delta $,
which is determined from the fitting ($\hbar \Omega $ at $\varepsilon _{1}=0$%
), can be varied by changing the gate voltage $V_{C}$ on the central gate $%
G_{C}$ as shown in Fig. 2(d). Although decoherence from first-order
tunneling processes is eliminated during manipulation, other decoherence
sources are significant in our measurement.

\begin{figure}[t]
\epsfxsize=3.3in \epsfbox{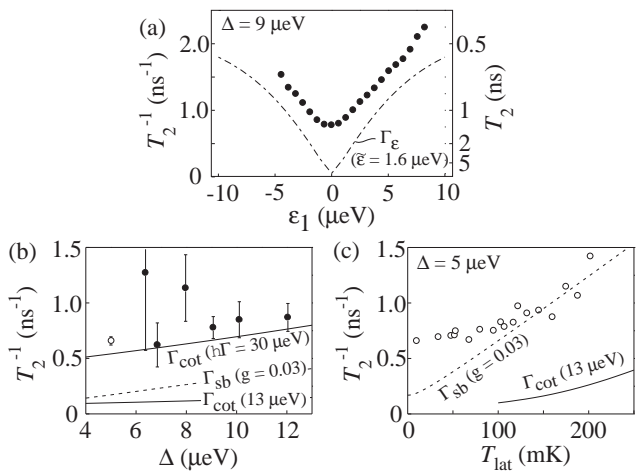} 
\caption{Decoherence rate, $T_{2}^{-1}$, of the qubit. (a) The energy offset
($\protect\varepsilon _{1}$) dependence. The dash-dotted line shows the
decoherence rate, $\Gamma _{\protect\varepsilon }$, due to the fluctuation $%
\tilde{\protect\varepsilon}=$ 1.6 $\protect\mu $eV. (b) The coupling energy (%
$\Delta $) dependence. (c) The lattice temperature ($T_{lat}$) dependence.
The solid (open) circles were measured with $\hbar \Gamma _{L}\sim \hbar
\Gamma _{R}$ $\sim $ 30 $\protect\mu $eV in sample I ($\hbar \Gamma _{L}\sim
\hbar \Gamma _{R}$ $\sim $ 13 $\protect\mu $eV in sample II). The
decoherence rates calculated from cotunneling ($\Gamma _{cot}$) and
spin-boson model ($\Gamma _{sb}$) are shown by solid and dashed lines,
respectively.}
\end{figure}

Firstly, background charge fluctuations and noise in the gate voltages
affect $\varepsilon $ and $\Delta $, which change the oscillation frequency $%
\Omega $ and dephase the system \cite{ItakuraPRB,NakamuraPRL}. The
fluctuation, $\tilde{\varepsilon}$, of $\varepsilon $ in our sample ranges
between 1.6 $\mu $eV, which is estimated from the fluctuation of $I$ in a
frequency range 0.1 - 5 Hz, and 3 $\mu $eV, which is the narrowest linewidth
of the resonant peak we obtained in sample I in the weak coupling limit ($%
\Delta <$ 1 $\mu $eV) \cite{FujisawaScience}. The corresponding decoherence
rate, $\Gamma _{\varepsilon }=|d\Omega /d\varepsilon |\tilde{\varepsilon}$
to the lowest order, for $\tilde{\varepsilon}=$ 1.6 $\mu $eV is shown by a
solid line in Fig. 3(a). This qualitatively explains the large decoherence
rate at $\varepsilon _{1}\neq 0$, where the system is sensitive to $\tilde{%
\varepsilon}$. However, the decoherence rate at $\varepsilon _{1}=0$ cannot
be explained with this model, and should be dominated by other mechanisms.

Secondly, we consider cotunneling effects. Although the first-order
tunneling processes are prohibited during manipulation, higher-order
tunneling (cotunneling) processes can occur because relatively high $\Gamma
_{L}$ and $\Gamma _{R}$ were chosen for efficient initialization. For
simplicity, we only estimate one of the cotunneling processes, which
scatters the electron from the anti-bonding state to the bonding state
(eigenstates of the qubit), from second order Fermi's golden rule. This
gives a transition rate $\Gamma _{cot}=(8/h)\Delta (\hbar \Gamma )^{2}/U^{2}$
at $\varepsilon _{1}=0$, $V_{sd}=0$ and zero temperature when the barrier is
symmetric ($\Gamma =\Gamma _{L}=\Gamma _{R}$) \cite{Eto}. $\Gamma _{cot}$
shown by solid lines in Figs. 3(b) and 3(c) actually includes thermal
broadening in the source and drain [$T_{elec}=$ 100 mK is assumed in Fig.
3(b)]. Although we cannot determine the parameters precisely, $\Gamma _{cot}$
is comparable to the observed $T_{2}^{-1}$. We believe that the cotunneling
effect is significant in our measurement but can be easily diminished by
choosing smaller $\Gamma $ and by making the interdot electrostatic coupling
energy $U$ larger.

Lastly, we discuss electron-phonon interactions, which is an intrinsic
decoherence mechanism in semiconductor QDs \cite%
{FujisawaScience,FujisawaNature}. Spontaneous acoustic phonon emission
remains even at zero temperature and causes the inelastic tunneling between
the two states \cite{FujisawaScience}. The phonon emission rate estimated
from the inelastic current or from the fitting with Eq. \ref{EqNp} is $%
\Gamma _{i}$ $\sim $ (4 - 20 ns)$^{-1}$, which depends on $\Delta $, at $%
\varepsilon =$ -30 $\mu $eV. $\Gamma _{i}$ of our interest at $\varepsilon =$
0 should be faster because of the spatial overlap of the eigenstates, and
may be comparable to the observed $T_{2}^{-1}$. By assuming Ohmic spectral
density for simplicity, the spin-boson model predicts the decoherence rate $%
\Gamma _{sb}=\frac{\pi }{4}g\Delta \coth (\Delta /2k_{B}T_{lat})$ for $%
\varepsilon =$ 0, where $g=$ 0.03 is the dimensionless coupling constant
that was chosen to fit with the temperature dependence data [see dashed line
in Fig. 3(c)] \cite{BrandesPRB,Leggett}. This $g$ is a reasonable value to
explain the inelastic current \cite{BrandesPRB}, and thus phonon emission
seems to be significant in our system.

Therefore, the qubit is strongly influenced by low-frequency fluctuation
when $|\varepsilon |\gtrsim \Delta $, cotunneling at high tunneling rates,
and acoustic phonons at high temperature. The resonances $\alpha $ and $%
\beta $ actually involve excited states in the right dot, and the relaxation
to the ground state should also cause decoherence. Other mechanisms, such as
the fluctuation of $\Delta $ and the electromagnetic environment, may have
to be considered to fully understand the decoherence. It should be noted
that the quality of the coherent oscillation was improved by reducing
high-frequency noise from the gate voltages and the coaxial cable. The
remaining noise may also contribute to the decoherence. We hope that some
decoherence effects can be reduced by further studies.

In summary, we have successfully manipulated the artificial qubit in a
double quantum dot. Coherent oscillations are observed for several
combinations of ground and excited states. In the present experiments, there
was no condition where two kinds of oscillations coincided, indicating that
the two-level system is still a good approximation. However, application of
a two-step voltage pulse, which consecutively adjusts the system at two
resonances ($\alpha $ and $\beta $ for instance) in a short time, would mix
three quantum states coherently. Therefore, DQDs are promising for studying
multi-level coherency\cite{BrandesPRL3level}, and the experiments can be
extended to electron-spin manipulations and two-qubit operations \cite%
{PashkinNature2qubit}.

We thank T. Brandes, T. Itakura, Y. Nakamura, K. Takashina, Y. Tokura, and
W. G. van der Wiel for their stimulating discussions and help.

\end{document}